\documentclass{aa}
\usepackage[varg]{txfonts}
\usepackage{amsmath}
\usepackage{amssymb}
\usepackage{graphicx}
\usepackage{inputenc}

\def\f#1   {Fig.~\ref{#1}}
\def\s#1   {Sec.~\ref{#1}}
\def\tab#1   {Tab.~\ref{#1}}
\def\eq#1   {Eq.~\ref{#1}}
\def\t#1   {Tab.~\ref{#1}}
\newcommand{\smo}{Smol\v{c}i\'{c} }

\title{The XXL Survey. XLI. Radio AGN luminosity functions based on the GMRT $610 \ \mathrm{MHz}$ continuum observations}
\titlerunning{The XXL Survey. XX}
\author{B. \v{S}laus\inst{1}
        \thanks{\emph{bslaus@phy.hr}
                University of Zagreb, Physics Department, Bijeni\v{c}ka cesta 32,
                10002 Zagreb, Croatia},
         V. \smo\inst{1}, M. Novak\inst{2}, S. Fotopoulou\inst{3,4}, P. Ciliegi\inst{5}, N. Jurlin\inst{6,7}, L. Ceraj\inst{1}, K. Tisani\'{c}\inst{1}, M. Birkinshaw\inst{8}, M. Bremer\inst{8}, L. Chiappetti\inst{9}, C. Horellou\inst{10},
         M. Huynh\inst{11, 12}, H. Intema\inst{13}, K. Kolokythas\inst{14}, M. Pierre\inst{15}, S. Raychaudhury\inst{16}, H. Rottgering\inst{13}  
}
\authorrunning{B. \v{S}laus}
\institute{Department of Physics, Faculty of Science, University of Zagreb,
Bijeni\v{c}ka cesta 32, 10000 Zagreb, Croatia   \and
Max Planck Institut f{\"u}r Astronomie, K{\"o}nigstuhl 17, 69117 Heidelberg, Germany\and
Center for Extragalactic Astronomy, Department of Physics, Durham University, South Road, Durham DH1 3LE, UK\and
Department of Astronomy, University of Geneva, ch. d’Ecogia 16, 1290 Versoix, Switzerland \and
INAF - Osservatorio Astronomico di Bologna, P. Gobetti 93/3, 40129 Bologna, Italy \and
ASTRON, the Netherlands Institute for Radio Astronomy, Postbus 2,7990 AA, Dwingeloo, The Netherlands \and
Kapteyn Astronomical Institute, University of Groningen, P.O. Box 800, 9700 AV Groningen, The Netherlands \and
H.H. Wills Physics Laboratory, University of Bristol, Tyndall Avenue, Bristol BS8 1TL, U.K. \and INAF, IASF Milano, via Corti 12, 20133 Milano, Italy\and 
Chalmers University of Technology, Dept. of Space, Earth and Environment, Onsala Space Observatory, 439 92 Onsala, Sweden \and
CSIRO Astronomy and Space Science, PO Box 1130, Bentley WA 6102, Australia  \and 
International Centre for Radio Astronomy Research (ICRAR), M468, The University of Western Australia, 35 Stirling Highway, Crawley, WA 6009, Australia \and
Leiden Observatory, Leiden University, Niels Bohrweg 2, 2333 CA, Leiden, The Netherlands \and   Centre for Space Research, North-West University, Potchefstroom 2520, South Africa \and AIM, CEA, CNRS, Universit\'e Paris-Saclay, Universit\'e Paris Diderot, Sorbonne Paris Cit\'e, F-91191 Gif-sur-Yvette,France  \and
Inter-University Centre for Astronomy and Astrophysics Ganeshkhind, Post Bag 4, Pun\'e 411 007, INDIA}

\begin{document}

\date{Received ?? Accepted ??}          
\abstract{We study the space density evolution of active galactic nuclei (AGN) using the $610 \ \mathrm{MHz}$ radio survey of the XXL-North field, performed with the Giant Metrewave Radio Telescope (GMRT). The survey covers an area of $30.4 \ \mathrm{deg}^2$, with a beamsize of $6.5 \ \mathrm{arcsec}$. The survey is divided into two parts, one covering an area of $11.9 \ \mathrm{deg}^2$ with $1 \sigma$ rms noise of $200\  \mathrm{\mu Jy \ beam^{-1}}$ and the other spanning $18.5 \ \mathrm{deg}^2$ with rms noise of $45\  \mathrm{\mu Jy \ beam^{-1}}$. We extracted the catalog of radio components above $7 \sigma$. The catalog was cross-matched with a multi-wavelength catalog of the XXL-North field (covering about $80 \%$ of the radio XXL-North field) using a likelihood ratio method, which determines the counterparts based on  their positions and their optical properties. The multi-component sources were matched visually with the aid of a computer code: Multi-Catalog Visual Cross-Matching (MCVCM). A flux density cut above $1\ \mathrm{mJy}$ selects AGN hosts with a high purity in terms of star formation  contamination based on the available source counts. After cross-matching and elimination of observational biases arising from survey incompletenesses, the number of remaining sources was $1150$. We constructed the rest-frame $1.4 \ \mathrm{GHz}$ radio luminosity functions of these sources using the maximum volume method. This survey allows us to probe luminosities of $ 23 \lesssim \log(L_{1.4 \ \mathrm{GHz}}[\mathrm{W/Hz}]) \lesssim 28$ up to redshifts of $z \approx 2.1$. Our results
are consistent with the results from the literature in which AGN are comprised of two differently evolving populations, where the high luminosity end of the luminosity functions evolves more strongly than the low-luminosity end.   }             

\maketitle      
\makeatother

\section{Introduction\label{sec:intro}}
\label{sec:Int}
It is now widely accepted that the evolution of AGN is closely related to the evolution of their host galaxies by a process called AGN feedback (e.g., \citealt{Heckman_Best2014}). Indirect proof of this connection can be deduced from the correlations between the masses of the central supermassive black hole and the properties of the host galaxies, for instance, the stellar velocity dispersion, the stellar mass of the bulge, or the bulge luminosity (\citealt{Magorrian1998}, \citealt{Ferrarese_Merritt2000}, \citealt{Gebhardt2000}, \citealt{Graham2011},  \citealt{Sani2011}, \citealt{Beifiori2012},  \citealt{McConnell_Ma2013}). A more direct proof for the importance of AGN feedback comes from the observation of galactic winds (e.g., \citealt{Nesvadba2008}, \citealt{Feruglio2010}, \citealt{Veilleux2013}, \citealt{Tombesi2015}) and X-ray cavities in groups and clusters of galaxies (\citealt{Clarke1997}, \citealt{Rafferty2006}, \citealt{McNamaraNulsen2007}, \citealt{Fabian2012}, \citealt{Nawaz2014}, \citealt{Kolokythas2015}). Furthermore, AGN feedback has become an essential element of state-of-the-art models of galaxy evolution  (e.g., \citealt{Croton2016}, \citealt{Harrison2018}). However, the mechanism of AGN feedback is not fully understood (e.g., \citealt{Cattaneo2009}, \citealt{Naab2017}). A useful tool to help understand these mechanisms and the timescales at which they are present is to study the evolution of radio luminosity functions (e.g., \citealt{Smolcic2009}, \citealt{Rigby2015},   \citealt{Pracy2016}, \citealt{Novak2018}). 

In order to disentangle the physical processes governing  AGN evolution, accretion onto the central supermassive black hole, and the feedback mechanism, prior studies  classify their radio sources into a number of distinct subsets. Concentrating on the underlying physics, studies generally suggest two fundamentally distinct populations. The first population consists of radiatively efficient AGN for which the accretion of cold gas onto the central black hole occurs at high Eddington ratios, $\lambda_{Edd}$, of $1\%$ to $10\%$ (\citealt{Heckman_Best2014}, \citealt{Smolcic2017a}, \citealt{Padovani2017}). This population is the one corresponding to the unified model of AGN widely present in the literature (e.g., \citealt{UrryPadovani1995}, \citealt{Netzer2015}). The second population is the radiatively inefficient population in which the accretion at lower Eddington ratios, typically $\lambda_{Edd} \lesssim 1\% $, is fueled by the hot intergalactic medium. This population is more prone to developing collimated jets (\citealt{Heckman_Best2014}). The different accretion efficiencies of the two populations seem to result from the differing physics between the optically thick geometrically thin disk accretion flow, which is radiatively efficient (\citealt{Shakura1973}), and the geometrically thick optically thin accretion flow (\citealt{Narayan1998}), as suggested by a number of studies (e.g., \citealt{Hardcastle2007}, \citealt{Heckman_Best2014}). Since the radiatively efficient mode exhibit emission lines in the optical spectra (due to the photoionization by the luminous disk), it is associated with high-excitation AGN. Depending on the strength of these lines the radio population is also often divided into high-excitation radio galaxies (HERGs) and low-excitation radio galaxies (LERGs) (\citealt{BestHeckman2012}).  Studies of the HERG and LERG radio luminosity functions in the local universe found that LERGs are the dominant population at luminosities below $L_{1.4 \ \mathrm{GHz}} \approx 10^{26} \ \mathrm{W \ Hz^{-1}}$, while HERGs dominate at the highest luminosities (\citealt{Pracy2016}, \citealt{BestHeckman2012}). The literature also suggests that AGN space density evolution is dependent on radio luminosity (e.g., \citealt{Smolcic2009}, \citealt{Willott2001}, \citealt{Waddington2001}, \citealt{Rigby2011},  \citealt{McAlpine2013}). It has been shown that the space density of the high-luminosity population evolves strongly with redshift up to $z\approx 2$. (\citealt{Dunlop1990}, \citealt{Willott2001}, \citealt{Pracy2016}), while the low-luminosity population exhibits little evolution (\citealt{Clewley2004}, \citealt{Smolcic2009}). The different evolution may be related to the different accretion modes.

Unlike optical surveys, radio observations are not affected by dust attenuation from the interstellar medium or absorbed by the Earth's atmosphere. Since high-luminosity sources are rare, in order to observe more of these sources, the area of observation must be large. The sensitivity of the observations, on the other hand, is the limiting factor concerning the observed redshifts. Here we present the radio luminosity functions of AGN within the XXL-North field, at $610 \ \mathrm{MHz}$ (\citealt{Pierre2016}, \citealt{SmolcicSlaus2018}). The observations cover a wide area ($30.4 \ \mathrm{deg}^2$) at high sensitivity (up to $45\  \mathrm{\mu Jy \ beam^{-1}}$) to constrain the evolution of the intermediate radio-luminosity population ($ 23 \lesssim \log(L_{1.4 \ \mathrm{GHz}}[\mathrm{W/Hz}]) \lesssim 28$) out to $z \approx 2$ at high sensitivity. 

The paper is organized as follows. In Sect. \ref{sec:Data} we describe the radio data and the corresponding multi-wavelength identifications of radio sources. In Sect. \ref{sec:CrCo} we describe the process of cross-matching via the likelihood ratio method. Section \ref{sec:LF} describes the creation of the luminosity functions, while Sect. \ref{sec:LFRes} presents the results and compares them with the literature. Results are discussed in Sect. \ref{sec:Dis}, while the summary and conclusion are given in Sect. \ref{sec:Sum}. Throughout this paper we use a cosmology defined with $H_0 = 70 \ \mathrm{km s^{-1} Mpc ^{-1}} $, $\Omega_m = 0.3$, and $\Omega_{\Lambda} = 0.7$. The spectral index, $\alpha$, was defined using the convention in which the radio emission is described as a power law, $S_{\nu} \propto \nu^{\alpha}$, where $\nu$ denotes the frequency, while $S_{\nu}$ is the flux density. We also use the $AB$ magnitude system.

\section{Data}
\label{sec:Data}
\subsection{Radio data}
\label{sec:radio}
The radio observations of the XXL-North field were performed at $610\  \mathrm{MHz}$ with the Giant Metrewave Radio Telescope (GMRT). The final mosaic of $79$ pointings encompasses an area of $30.4 \ \mathrm{deg}^2$. Observations of the inner $36$ pointings (the XMM-Large Scale Structure, XMM-LSS field) were taken from an earlier study by \citet{Tasse2007}, and re-reduced for the purposes of this study. They encompass an area of $11.9 \ \mathrm{deg}^2$ and reach a mean rms of $200\  \mathrm{\mu Jy \ beam^{-1}}$. The remaining $18.5 \ \mathrm{deg}^2$, observed by  \citet{SmolcicSlaus2018} (hereafter XXL Paper XXIX) have a mean rms of $45\  \mathrm{\mu Jy \ beam^{-1}}$. The FWHM of the synthesized beam across the entire mosaic is $6.5\  \mathrm{arcsec}$. The data reduction and imaging were performed using the Source Peeling and Atmospheric Modeling (SPAM) pipeline (\citealt{Intema2009}, \citealt{Intema2017}). Source extraction, performed with the PyBDSF\footnote{https://www.astron.nl/citt/pybdsf/} software (\citealt{Mohan_Rafferty2015}), resulted in the identification of $5434$ sources with a conservative signal-to-noise ratio of $S/N\geq 7$.

A pre-selection of possible multi-component sources was performed via an automatic method following the methods in \citet{Tasse2006}. All sources whose separations were within $60\  \mathrm{arcsec}$ were tagged in a separate catalog column. An additional flux limit of $S_{610\ \mathrm{MHz}} > 1.4\  \mathrm{mJy} $ was introduced in the outer parts of the field, justified by the size--flux relation for radio sources, where larger sources emit more flux (\citealt{Bondi2003}). For the final classification of multi-component sources see Sect. \ref{sec:MC}. The radio catalog also contains spectral indices estimated using the NRAO Very Large Array Sky Survey (NVSS; \citealt{Condon1998}). For further details concerning the GMRT radio observations and the corresponding catalog we refer the reader to Paper XXIX.

\subsection{Multi-wavelength catalog}
\label{MultiCat}
The XXL-North field has been surveyed in a wide range of different bands (from radio to X-ray). In the current paper we use only the subset of the catalog that has identifications in the Spitzer Infrared Array Camera (IRAC) Channel 1 band at $3.6 \ \mathrm{\mu m}$ (PI M. Bremer, limiting magnitude of 21.5 AB). This provides us with a catalog of uniform density and depth. The photometric redshifts of the IRAC-detected sources are obtained from the full multi-wavelength data (Fotopoulou in prep.).

\begin{figure}
\includegraphics[width=0.45\textwidth]{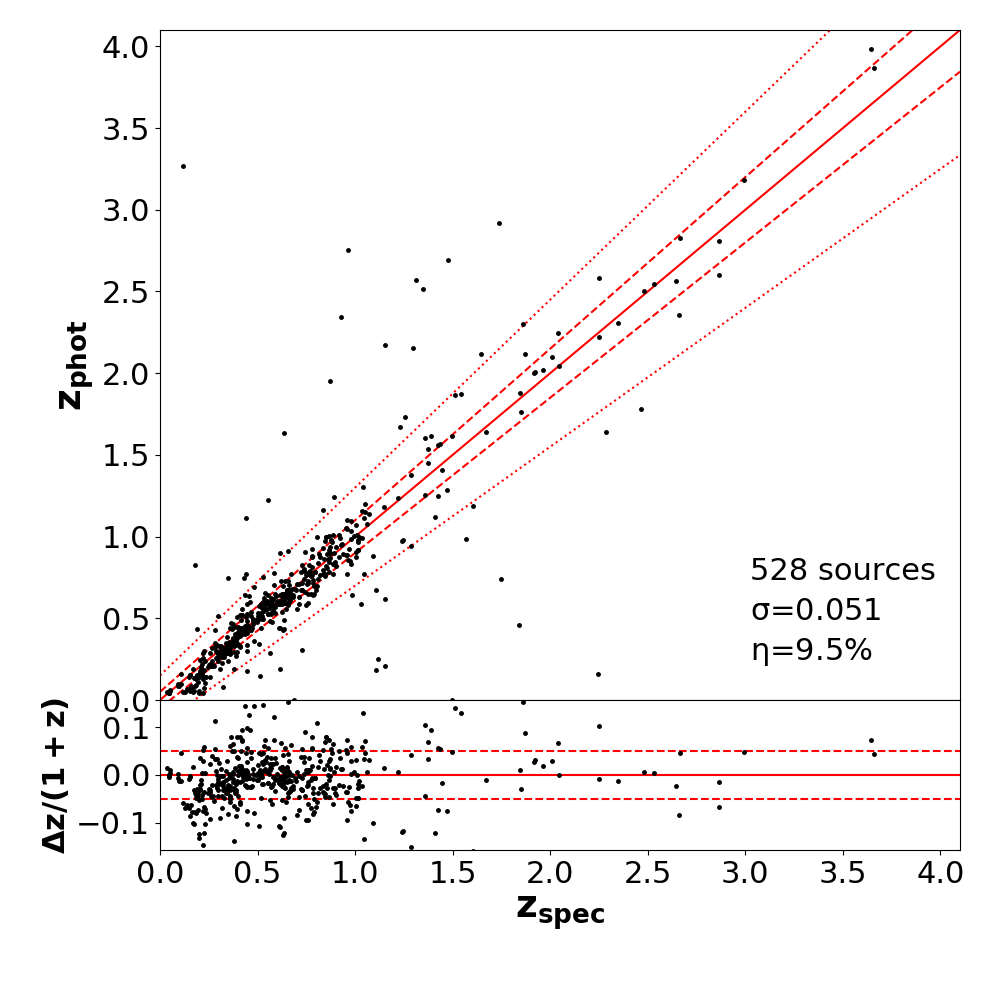}
\caption{Comparison between the spectroscopic ($z_{\mathrm{spec}}$) and the photometric ($z_{\mathrm{phot}}$) redshifts for $528$ sources with good quality spectra. For the definition of accuracy $\sigma$ and the percentage of catastrophic outliers $\eta$, see the text. The bottom panel shows the renormalized accuracy, defined as denoted in the figure.}
\label{fig:Spec-Photo}
\end{figure}

The wealth of data allowed  the creation of a multi-wavelength catalog of the XXL-North field and the calculation of photometric redshifts. The creation of the photometric catalog can be found in \citet{Fotopoulou2016} (XXL Paper VI), while the photometric redshift estimation method is described in detail in \citet{Fotopoulou2018}. We cross-matched the photometric catalog with Sloan Digital Sky Survey Data Release 14 (SDSS DR14) and our database of spectroscopic follow-up redshift observations of the XXL survey (\citealt{Adami2018}, XXL Paper XX) and found $408$ and $120$ good quality spectra within $1 \ \mathrm{arcsec}$ from the the GMRT counterpart. Based on this spectroscopic sample, the photometric redshifts of the radio counterparts reach an accuracy of $\mathrm{\sigma_{NMAD}=0.05}$ with $\mathrm{\eta=9.5\%}$ catastrophic outliers\footnote{The accuracy is defined as $\sigma=1.48\frac{|z_{phot}-z_{spec}| }{ 1+z_{spec} }$ and the number of catastrophic outliers is the fraction of sources with $N[\frac{|z_{phot}-z_{spec}| }{ 1+z_{spec} }]>0.15$.}. The comparison between the photometric and the spectroscopic redshifts is shown in Figure \ref{fig:Spec-Photo}. The solid line shows the one-to-one relationship, while the dashed and dotted lines correspond to $\mathrm{z_{phot}=0.05\cdot(1+z_{spec})}$ and $\mathrm{z_{phot}=0.15\cdot(1+z_{spec}),}$ respectively.

\begin{figure}[htb]
\centering
\includegraphics[width=0.5\textwidth]{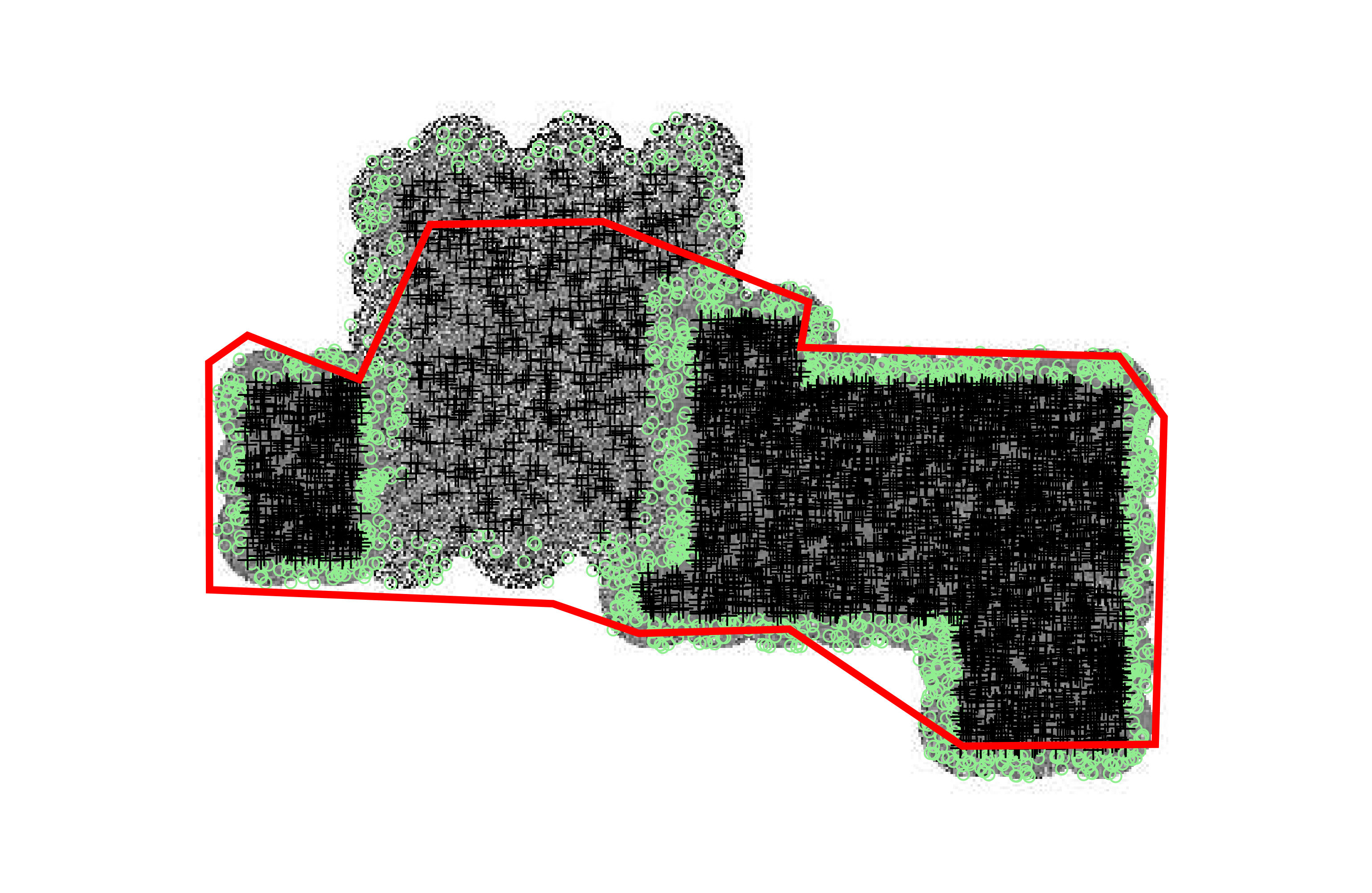}
\caption{Spitzer IRAC $3.6 \ \mathrm{\mu m}$ coverage of the XXL-North field. The gray map corresponds to the GMRT $610 \ \mathrm{MHz}$ mosaic. The red region denotes the IRAC data. The largest mismatch occurs in the inner part of the radio field, where the number of sources is lower. The sources in the radio catalog are denoted by green circles and black crosses. Sources marked by green symbols correspond to the noisy edges, as described in the text. }
\label{fig:XXLN-Coverage}
\end{figure}

As shown in Fig. \ref{fig:XXLN-Coverage}, the Spitzer IRAC $3.6 \ \mathrm{\mu m}$ map does not cover the area of our radio observations completely. The overlap of the radio data with the IRAC coverage is $8.0\ \mathrm{deg.}^2$ (roughly $67 \%$) for the inner part of the radio mosaic and $16.7\ \mathrm{deg.}^2$ (i.e., roughly $90 \%$) for the outer parts (or roughly $80 \%$ for the complete XXL-North field). The majority of sources lie in the outer (and deeper) part of the mosaic, which is well covered by the IRAC survey.

\section{Cross-matching of sources}
\label{sec:CrCo}
\subsection{Mean positional offsets}
\begin{figure}
\includegraphics[width=0.45\textwidth]{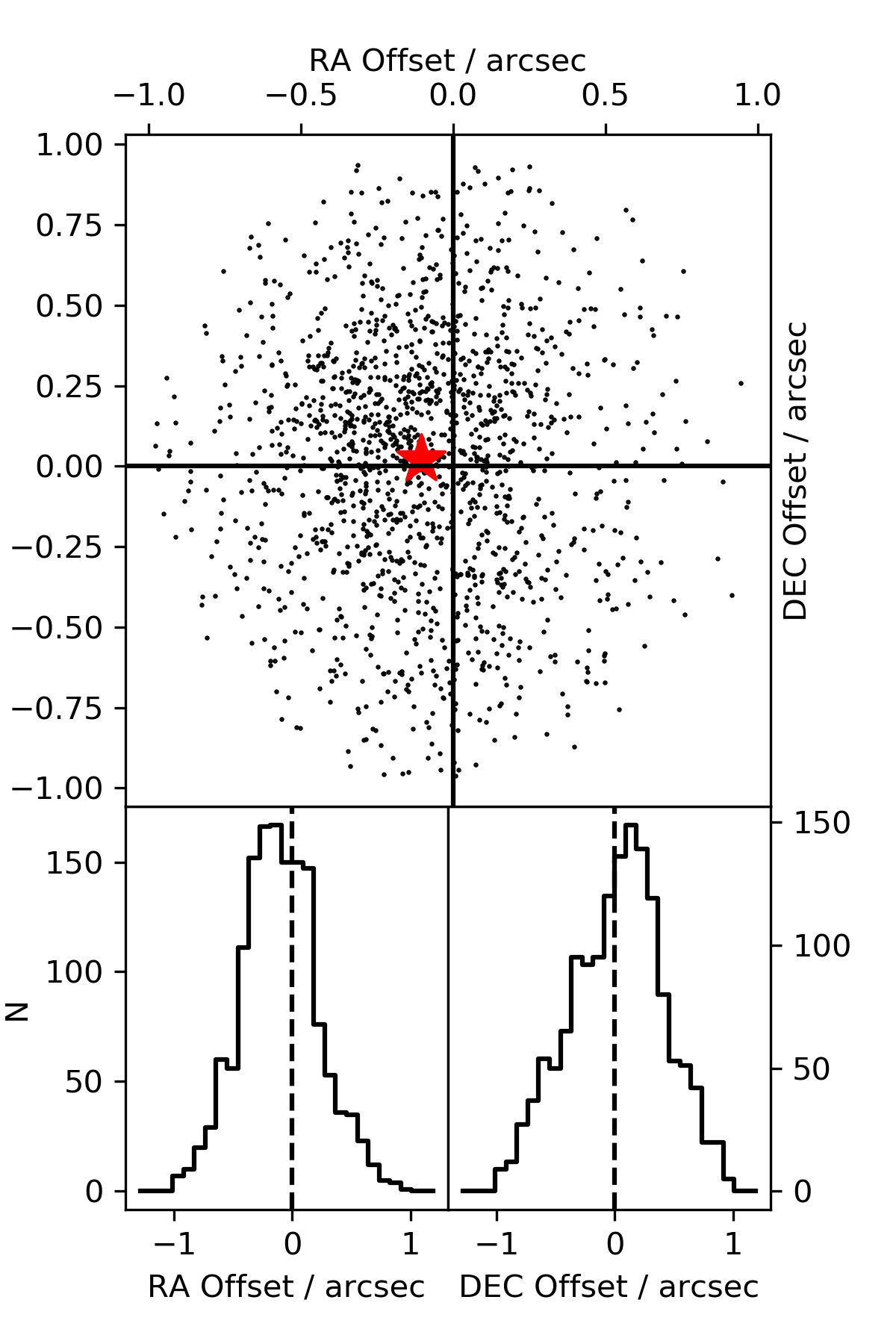}
\caption{Astrometric offsets between the GMRT and IRAC surveys for the outer part of the XXL-North field. The mean offset is denoted by a red star. The histograms in the bottom panels represent the distribution of offsets in the $RA$ and $DEC$ directions. The inner part of the XXL-North field produces a consistent plot.}
\label{fig:Astro-Offset}
\end{figure}

Prior to the cross-matching of sources, we assessed the mean systematic offset between the GMRT and IRAC source positions. We performed a simple match between the two fields based solely on the source positions, selecting only sources whose positional offset is $1 \ \mathrm{arcsec}$ or less. We show the positional differences between the two surveys in Figure \ref{fig:Astro-Offset}. To minimize the contribution from spurious counterparts we limited the GMRT radio sample to unresolved sources with signal-to-noise ratio of $S/N > 10$. The obtained matches, although highly incomplete, were considered very reliable. The mean positional offset in the $RA$ and $DEC$ coordinates between the GMRT and the IRAC positions of the matched sources are 
\begin{align} 
\overline{\Delta RA} = (0.02 \pm 0.03) \ \mathrm{arcsec},\ \\ \overline{\Delta DEC} = (0.07 \pm 0.02) \ \mathrm{arcsec}
\end{align}
for the inner (XMM-LSS) part of the GMRT mosaic, and
\begin{align} 
\overline{\Delta RA} = (0.104 \pm 0.008) \ \mathrm{arcsec},\ \\ \overline{\Delta DEC} = (0.02 \pm 0.01) \ \mathrm{arcsec}
\end{align}
for the rest of the XXL-North field. Although the offsets were not large, we eliminated them from further considerations by correcting the relative distances between the sources.

\subsection{Multi-component sources}
\label{sec:MC}
\begin{figure}
\includegraphics[width=0.45\textwidth]{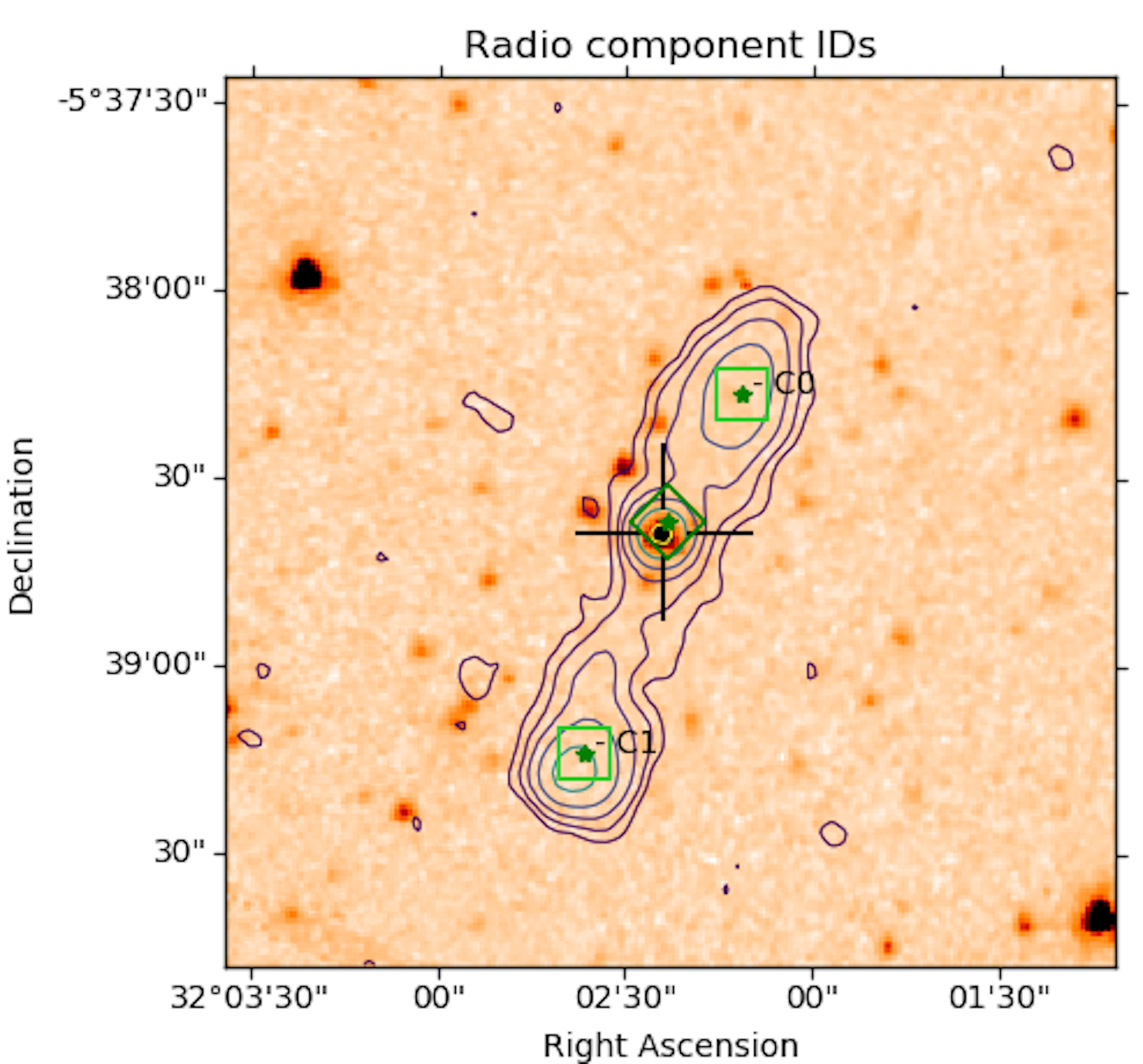}
\caption{Example of the image created by the MCVCM program. The radio contours (chosen as $2^n \times \mathrm{RMS},\ n=1,2,3...$) are overlaid on the IRAC image. The radio and IRAC selection is performed visually. The black crosshair denotes the IRAC counterpart position. The dark green rhomboid denotes the radio core position, while the light green squares denote the center of the radio lobes.}
\label{fig:MCVCM}
\end{figure}

Sources with complex morphologies might be recovered not as single objects, but as a collection of components due to a limited surface brightness sensitivity of radio surveys (\citealt{Schinnerer2004}, \citealt{Schinnerer2007}, \citealt{Smolcic2017a}). This will introduce errors during the cross-matching with other surveys and affect the luminosity functions. We performed a multi-component source classification by visually inspecting sources pre-selected via the automatic method described in Sect. \ref{sec:radio}. In order to expedite the process we used the publicly available MCVCM package\footnote{\url{https://github.com/kasekun/MCVCM}}, which simplifies the visual cross-matching, by producing IRAC $3.6 \ \mathrm{\mu m}$ images of the sources with overlaid radio contours. An example of the created images can be seen in Figure \ref{fig:MCVCM}. The radio core and lobe components as well as the infrared centroid were selected manually by visual inspection. Using this method, we classified $381$ components belonging to multi-component sources. The radio fluxes of these components were summed and their positions taken to be that of the IRAC (infrared detected) centroid source. The final number of multi-component sources was $157$. The sources matched with this method were excluded from further matching via the likelihood ratio method, described in the following sections.

The automatic method described in Sect. \ref{sec:radio}, although generally reliable, missed six conspicuous multi-component galaxies due to the size of these sources (more than $60\  \mathrm{arcsec}$). These six large galaxies were therefore re-matched manually (using again the MCVCM program). In the final catalog from Paper XXIX they are denoted by the following names: XXL-GMRT J$023357.0-050753$, XXL-GMRT J$023110.7-053314$, XXL-GMRT J$021659.0-044837$, XXL-GMRT J$021003.1-052825$, XXL-GMRT J$020759.2-065019$, XXL-GMRT J$020354.8-041356$. Two of these galaxies (XXL-GMRT J$021003.1-052825$ and XXL-GMRT J$020354.8-041356$) were studied in detail by \citet{Horellou2018}, XXL Paper XXXIV.

\subsection{Likelihood ratio method}
For the cross-matching of single-component sources, between the GMRT XXL-North $610\  \mathrm{MHz}$ survey and the IRAC survey, we used the likelihood ratio ($LR$) method (\citealt{Sutherland_Saunders1992}; see also \citealt{deRuiter1977},  \citealt{Ciliegi2003}, \citealt{Brusa2007}, \citealt{Mainieri2008}, \citealt{Smith2011}, \citealt{Bonzini2012}, \citealt{McAlpine2012}, \citealt{Fleuren2012}, \citealt{Kim2012}). The $LR$ of each possible identification is defined as the probability between the source being a true counterpart and it being an unrelated background object. By assuming that the optical properties of the sources are independent of their positional offsets (\citealt{Sutherland_Saunders1992}, \citealt{Ciliegi2003}), the expression for $LR$ becomes
\begin{align} 
LR = \frac{f\left(r\right) q\left(m\right)}{n\left(m\right)},
\end{align}
where $f \left( r \right)$ is the probability distribution of the positional offsets between the surveys, $q \left( m \right)$ the expected magnitude distribution of true counterparts, and $n \left( m \right)$ the surface density of the unrelated background objects given as a function of magnitude. The magnitudes here correspond to the IRAC $3.6 \ \mathrm{\mu m}$ magnitudes of the possible counterparts of the radio sources.

\subsection{Estimation of $f \left( r \right)$}
For the radial probability distribution of positional offsets we used a Gaussian function (\citealt{Ciliegi2003}, \citealt{Smith2011}, \citealt{Bonzini2012}, \citealt{McAlpine2012}, \citealt{Fleuren2012}, \citealt{Kim2012})
\begin{align} 
f\left(r\right) = \frac{1}{\sqrt{2\pi\sigma^{2}}} \mathrm{exp} \left( -\frac{r^{2}}{2\sigma^{2}}\right),
\end{align}
where $r$ denotes the separation between the GMRT $610 \ \mathrm{MHz}$ and the IRAC $3.6 \ \mathrm{\mu m}$ source positions. The standard deviation of the distribution $\sigma$ is obtained from the positional uncertainties of both surveys $\sigma_{GMRT}$ and $\sigma_{IRAC}$. Following \citet{Ciliegi2003} we defined the standard deviation as
\begin{align} 
\sigma = \sqrt{\sigma_{IRAC}^{2} + \sigma_{GMRT}^{2}} .
\end{align}
For the GMRT data we used the positional errors listed in the radio source catalog  provided by PyBDSF (with a mean value of around $0.2 \ \mathrm{arcsec}$ for both parts of the field and both coordinates). The IRAC positional errors were calculated from the full width at half maximum (FWHM) of the IRAC $3.6 \ \mathrm{\mu m}$ beam and the signal-to-noise ratio of each source ($S/N$), following \citet{Ivison2007} and \citet{Furlanetto2018} as 
\begin{align} 
\sigma = 0.6 \frac{FWHM}{S/N}.
\end{align}
Furthermore, the errors were not allowed to be smaller than $0.2 \ \mathrm{arcsec}$ (roughly one-third of the mean positional error which was about $0.6 \ \mathrm{arcsec}$) to account for the minimum positional uncertainty as discussed in \citet{Smith2011}. In order to account for possible anisotropies in the positional errors, we calculated $\sigma$ separately in the RA and DEC directions. The final standard deviation, used in the probability distribution of positional offsets, is the mean value between the two.  Although $f \left( r \right)$ is normalized to unity for radii spanning to infinity, in practice a fixed value of maximum radius is set during the cross-matching. The maximum allowed separation $R$ is called the matching radius. In this paper we used a matching radius of $4\ \mathrm{arcsec}$.

\subsection{Estimation of $n \left( m \right)$}

The background source density as a function of magnitude $n \left( m \right)$ was obtained by normalizing the magnitude distribution of the complete IRAC $3.6 \ \mathrm{\mu m}$ catalog by the area of the IRAC $3.6 \ \mathrm{\mu m}$ survey (\citealt{Smith2011}, \citealt{Furlanetto2018}). Our main assumption here is that the shape of the background magnitude distribution is equal to the shape of the magnitude distribution of the complete IRAC catalog, which is sufficiently accurate if the number of real identifications is much smaller than the total number of IRAC sources.

Since the number of radio sources is small we can assume that the circles defined by the matching radius $R$ do not overlap. The average number of unrelated background objects within the area defined by the matching radius $R$ around each radio source is given then by
\begin{align} 
\label{eq_n}
false(m) = n \left( m \right) \cdot N_{Radio} \cdot  \pi R^{2} ,
\end{align}
where $N_{Radio}$ is the number of radio sources, corresponding only to the sources within the area covered by both GMRT XXL-North $610 \ \mathrm{MHz}$ and IRAC surveys (roughly $80 \%$ of the area of the mosaic, as described in Sect. \ref{MultiCat}).

\subsection{Estimation of $q \left( m \right)$}

To estimate the expected distribution of true counterparts, $q \left( m \right)$, we created the magnitude distribution of the total number of possible counterparts within the matching radius $R=4\ \mathrm{arcsec}$, $total\left( m \right)$. This distribution also contains the false counterpart identifications arising from the unrelated background sources (eq. \ref{eq_n}). Following \citet{Ciliegi2003}, we constructed a new magnitude distribution, $real\left( m \right)$, which is the difference between the total and the background distributions:
\begin{align} 
real\left( m \right) = total\left( m \right) - false(m) .
\end{align}
This excess of sources compared to the background distribution represents the expected real identifications. The resulting distribution was further normalized as
\begin{align} 
q \left( m \right) = \frac{real\left( m \right)}{\sum_{m} real\left( m \right)} \cdot Q ,
\end{align}
where the sum in the denominator sums the $real\left( m \right)$ distribution over magnitudes. The $Q$ factor is the fraction of true counterparts above the magnitude limit (\citealt{Smith2011}), i.e., a correction for the limiting magnitude of our observations. It was calculated by summing the $real\left( m \right)$ distribution and dividing it by the number of radio sources (in the intersection):
\begin{align} 
\label{eqQ}
Q = \frac{\sum_{m} real\left( m \right)}{N_{Radio}} .
\end{align}
The value of $Q$ is $0.62$ for the outer part of the field and $0.55$ for the inner. It should be noted, however, that the value of $Q$ does not affect the results of the cross-matching significantly, as already noted by earlier studies (\citealt{Ciliegi2003},  \citealt{Franceschini2006}, \citealt{Fadda2006}, \citealt{Mainieri2008}).

\begin{figure}
\includegraphics[width=0.45\textwidth]{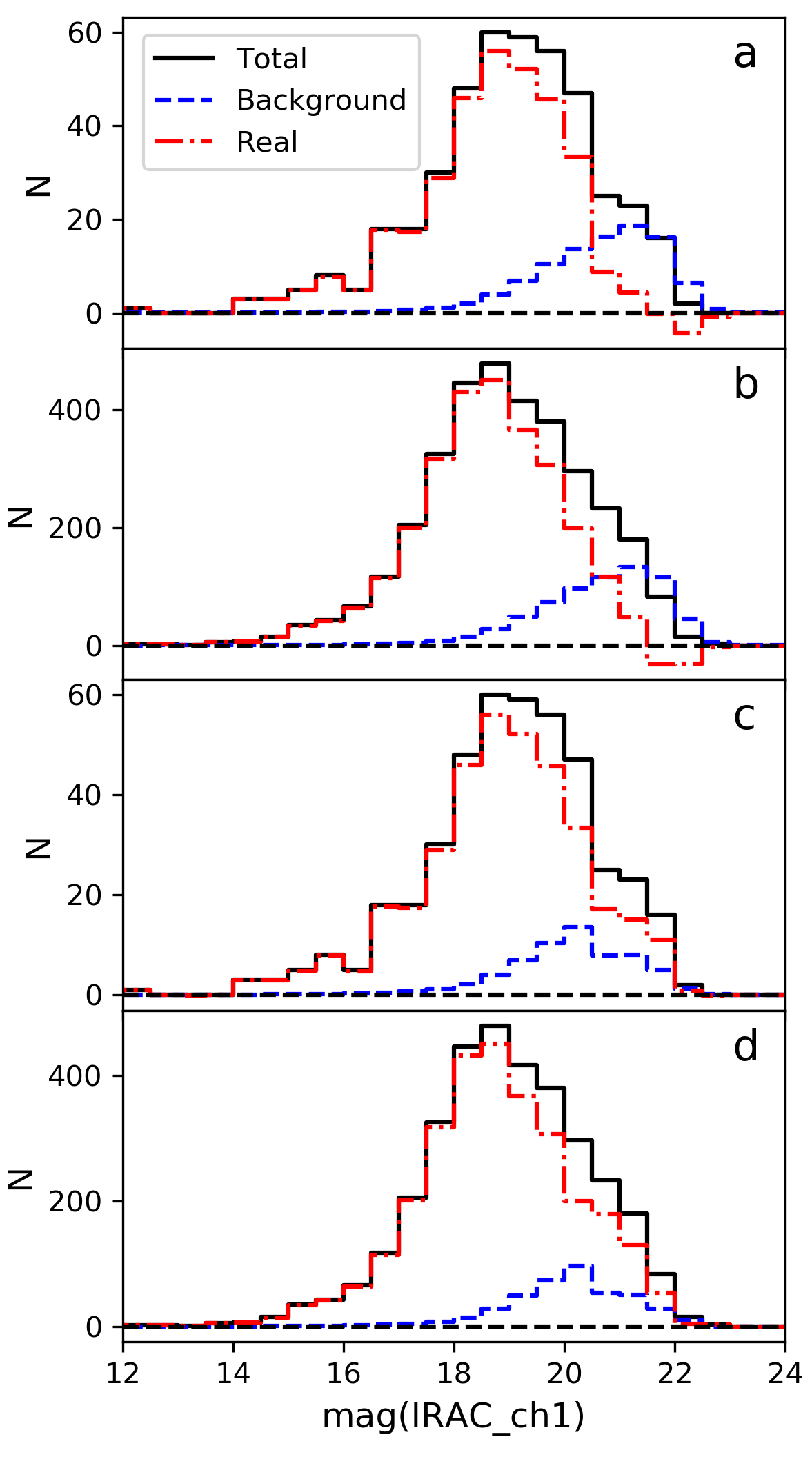}
\caption{Magnitude distribution of sources during the matching process. The black, blue, and red lines denote the total, background, and real sources, respectively, as described in the text. The two upper histograms (panels $a$ and $b$, for the inner and outer part of the field, respectively) correspond to the match where the blocking effect is present. Correction for blocking effects mitigates the issue of negative counts. The two bottom histograms (panels $c$ and $d$, for the inner and outer part of the field, respectively) are the magnitude distributions after the blocking effect has been accounted for.   }
\label{fig:MagD}
\end{figure}

\subsection{Blocking effect}
A further complication arises because of the tendency of radio sources to have bright counterparts (\citealt{Ciliegi2018}, hereafter XXL Paper XXVI). Some faint IRAC sources around these bright infrared counterparts remain undetected. Since the total number of counterparts is calculated within the matching radius around the radio positions, this leads to the underestimating of the $total\left( m \right)$ magnitude distribution. We call this effect the blocking effect, since the faint IRAC sources are blocked by the bright ones. On the other hand, the background density distribution $n\left( m \right)$, obtained from the complete IRAC catalog, is not significantly affected by this effect since the number of bright sources in the complete catalog is small. It follows that at faint magnitudes the $real\left( m \right)$ distribution becomes underestimated and even assumes unphysical negative values. The magnitude distributions are presented in Fig. \ref{fig:MagD}. This effect and the required correction have already been discussed by \citet{Brusa2007}, \citet{Smolcic2017}, and in Paper XXVI.  

In order to account for the missing sources we recalculated the magnitude distribution of the unrelated background sources following \citet{Brusa2007} and Paper XXVI. First we selected a random sample of $5000$ sources from the IRAC catalog that followed the same magnitude distribution as the $total\left( m \right)$ counterparts. We then used these sources as a mock radio catalog and re-counted the remaining IRAC sources in their vicinity (within $R=8\ \mathrm{arcsec}$). By re-normalizing this number to the number of radio sources and the correct matching radius with a factor
\begin{align} 
\frac{N_{Radio} \cdot \pi \cdot (4 \mathrm{arcsec})^{2}}{5000 \cdot \pi \cdot (8 \mathrm{arcsec})^{2}} ,
\end{align}
we were able to obtain the new estimate for the background magnitude distribution. The main advantage of the new background estimate was that, by definition, it included the blocking effect present within the IRAC catalog. In other words, the new background density is no longer overestimated compared to the number of total counterparts since it was also calculated around other bright IRAC sources. The resulting background distribution was consistent with the global one for bright magnitudes, but differed strongly for faint magnitudes. We therefore took the global background distribution at bright magnitudes down to a fixed limit of $m_{AB} = 20.5$ in IRAC magnitudes. At fainter magnitudes we used the new estimation (based on the mock radio catalog) of the background distribution described above, which resulted in a larger number of faint identifications considered real. The distributions estimated by this method are shown in Fig. \ref{fig:MagD}.

\subsection{Cross-matching results}
Following the literature (e.g., \citealt{Mainieri2008}, Paper XXVI), we limited the final catalog to sources with $LR > 0.2$. In addition to   the likelihood ratio, we can also define the reliability (e.g., \citealt{Franceschini2006}, \citealt{Fleuren2012}, \citealt{Butler2017}, hereafter XXL Paper XVIII) as
\begin{align} 
Rel_i = \frac{LR_i}{\sum_i LR_i + (1-Q)},
\end{align}
where $Q$ is given by equation \ref{eqQ}. In the case of multiple identifications with $LR>0.2$ we chose the counterpart with the largest reliability (\citealt{Mainieri2008}, Paper XVIII) resulting in $3336$ sources (see Table \ref{tab:1}). Finally, we excluded sources lying in the noisy edges of the radio map since the RMS noise was deemed too high (see Fig. \ref{fig:XXLN-Coverage}). The edges were defined manually, as described by Paper XXIX (see Fig. 5 from that paper for noise distribution). The final matched catalog, which also  includes  the visually matched sources, consists of $2467$ sources in the outer part of the field and $318$ in the inner ($2785$ in total, see Table \ref{tab:1}). This corresponds in total to roughly $60 \%$ of the sources in the intersection being matched. All of the matched sources have a reliable redshift estimation. Concentrating on only the area away from the noisy edges, the percentage of matches is around $67 \%$, which is in agreement with the literature for similar surveys (e.g., Paper XXVI).

\renewcommand{\labelitemi}{$-$}
\subsection{Source catalog description}
The results of the cross-matching performed within this work were compiled into a source catalog. The radio positions and the corresponding uncertainties were provided by the PyDBSF, and taken from an earlier catalog described in Paper XXIX and briefly discussed in Sect. \ref{sec:radio}. The flux densities come from the same radio catalog. The photometric redshift is obtained from the multi-wavelength counterpart catalog (Fotopoulou in prep.) described in Sect. \ref{MultiCat}. As already stated in Sect. \ref{sec:MC}, some of the sources were matched manually using the MCVCM package. These sources are present in the catalog, but are lacking some of the data (denoted by $-99.99$). The position of these sources is the position of the counterpart source and the integrated radio flux density is the sum of all the radio flux densities of the corresponding radio components. The columns are named as follows:
\begin{itemize}
   \item Name: Name of the radio source
   \item ID: Numeric identifier of the radio source
   \item RA: Right ascension of the radio source
   \item DEC: Declination of the radio source
   \item E\_RA: Uncertainty on the RA radio source position
   \item E\_DEC: Uncertainty on the DEC radio source position
   \item Peak\_flux: Peak radio flux density in $\mathrm{Jy/beam}$
   \item RMS: Local RMS in $\mathrm{Jy/beam}$ 
   \item Total\_flux: Integrated flux density of the radio source in $\mathrm{Jy}$
   \item E\_Total\_flux: Uncertainty on the integrated radio flux density   
   \item Alpha: The spectral index of the source.
   \item RA\_IRAC: Right ascension of the IRAC-detected counterpart
   \item DEC\_IRAC: Declination of the IRAC-detected counterpart
   \item Photo\_Z: Photometric redshift of the source
   \item LR: Likelihood ratio of the counterpart source as described in Sect. \ref{sec:CrCo} 
   \item Area\_Flag: Tag column denoting the inner (XMM-LSS) and outer part of the XXL-North field, described in Sect. \ref{sec:radio}. Zero denotes the inner part of the field.
   \item Edge\_Flag: Tag column denoting the sources lying in the noisy edges of the field. Zero denotes sources on the edge.   
   \item New\_Flag: Tag column denoting newly created multi-component sources matched manually with MCVCM. New sources: 1; new sources with names identical to the sources from the Paper XXIX catalog: 2.
\end{itemize}
The catalog is available as queryable database table XXL\_GMRT\_17\_ctpt
via the XXL Master Catalogue browser\footnote{\url{http://cosmosdb.iasf-milano.inaf.it/XXL}}. A copy will also
be deposited at the Centre de Donn\'ees astronomiques de Strasbourg
(CDS)\footnote{\url{http://cdsweb.ustrasbg.fr}}

\subsection{Missing counterparts: Comparison with the COSMOS data}
\label{IRAC_Corr}
\begin{figure}
\includegraphics[width=0.45\textwidth]{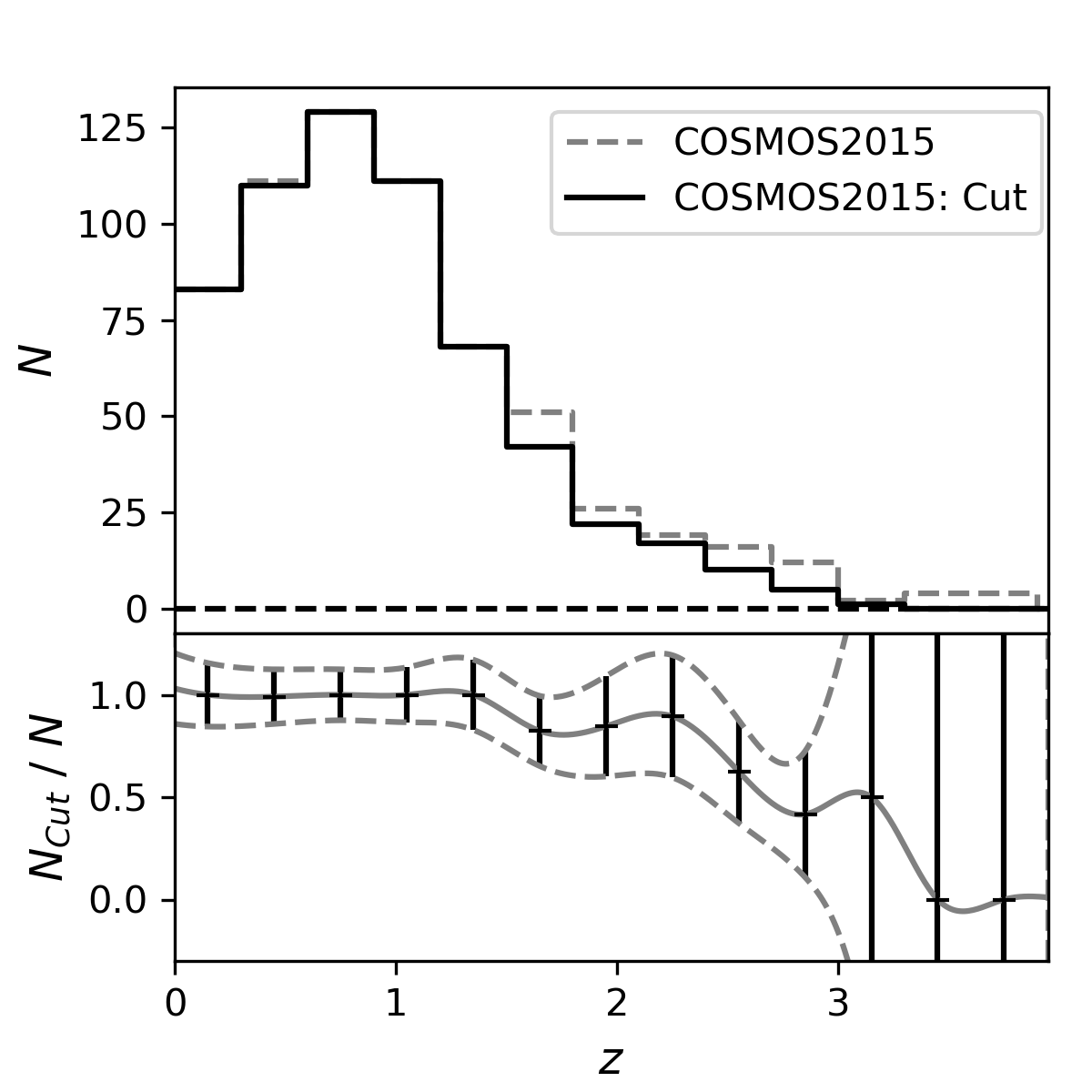}
\caption{Upper panel: Redshift histograms of the COSMOS2015 catalog with the radio cut (dashed gray line) and the histogram with an additional cut in the infrared flux corresponding to the IRAC detection limit of our survey (black line). Bottom panel: Ratio of these two histograms to the corresponding standard deviation. A cubic interpolation has been performed on both the data points and the error bars. }
\label{fig:CosmosCut}
\end{figure}

Since the IRAC data used in the cross-matching are of medium-depth ($m_{AB} =21.5$) it is necessary to assess the number of sources that are lost during the matching process and how this deficit of sources scales with redshift. To examine this problem, we used the deeper radio data from the VLA-COSMOS $3 \ \mathrm{GHz}$ Large Project  detected above a $5\sigma$ threshold of $11.5 \ \mathrm{\mu Jy}$ (\citealt{Smolcic2017}, b). \citet{Smolcic2017} cross-matched the $3 \ \mathrm{GHz}$ data with the multi-wavelength COSMOS2015 catalog (\citealt{Laigle2016}), which contains Channel 1 IRAC sources (as described by \citealt{Laigle2016}). The cross-matching of the $3 \ \mathrm{GHz}$ and IRAC data found counterparts for $\approx 93 \%$ of radio sources (see \citealt{Smolcic2017} for details).
We imposed a threshold in flux density on the COSMOS data equal to our radio detection limit ($350 \ \mathrm{\mu Jy}$ shifted to COSMOS frequencies by assuming a power law and a mean spectral index of $-0.7$) in order to mimic our radio data. This also ensures that this subsample of COSMOS sources is complete over all redshifts studied with our data ($0.1<z<2.1$; see Fig. 16. from \citealt{Smolcic2017a} or Fig. 1. from \citealt{Delvecchio2017} for details). We created the redshift histogram of this dataset. Then we examined the same dataset with an additional threshold corresponding to the IRAC detection limit of our survey, and re-created the redshift histogram. The comparison between these two histograms quantifies the sources lost during the cross-matching. The histograms detailing the redshift dependency of the comparison can be seen in Fig. \ref{fig:CosmosCut}. The bottom panel of Fig \ref{fig:CosmosCut} shows the ratio of the two distributions which we regarded as the necessary correction $C_{IRAC}(z)$. For the standard deviation of the histograms we assumed the Poissonian deviation, which scales with the number of sources as $\sqrt{N}$ except when the number of sources is lower than $N = 10$. In this situation we calculated the standard deviation as $N+\sqrt{N+0.75}$, following the approximation for the upper limit error bars from \citet{Gehrels1986}. Since the deviation in these bins is rather large and we needed only a rough approximation, we presumed the error bars were symmetrical. It can be seen that around redshift of $3$ the fraction drops to values of $\approx 0.5$ and the uncertainties become very large (i.e., comparable with the values of the fraction). We therefore used only $z \leq 2.1$ data for the further analysis.

\section{Radio luminosity functions of AGN}
\label{sec:LF}
In this section we describe the creation of the luminosity functions of our sample. The photometric redshifts were taken from the multi-wavelength catalog (see Sect. \ref{MultiCat}). In Sect. \ref{AGN_Fract} we describe the galaxy populations comprising our sample. In Sect. \ref{AGN_Cre} we describe the computation of the luminosity functions via the maximum volume method. The complete account of the required corrections is described in Sect. \ref{LFCorr} while details on the bin selection are given in Sect. \ref{BinSelection}.

\subsection{Galaxy populations}
\label{AGN_Fract}
\begin{figure}
\includegraphics[width=0.45\textwidth]{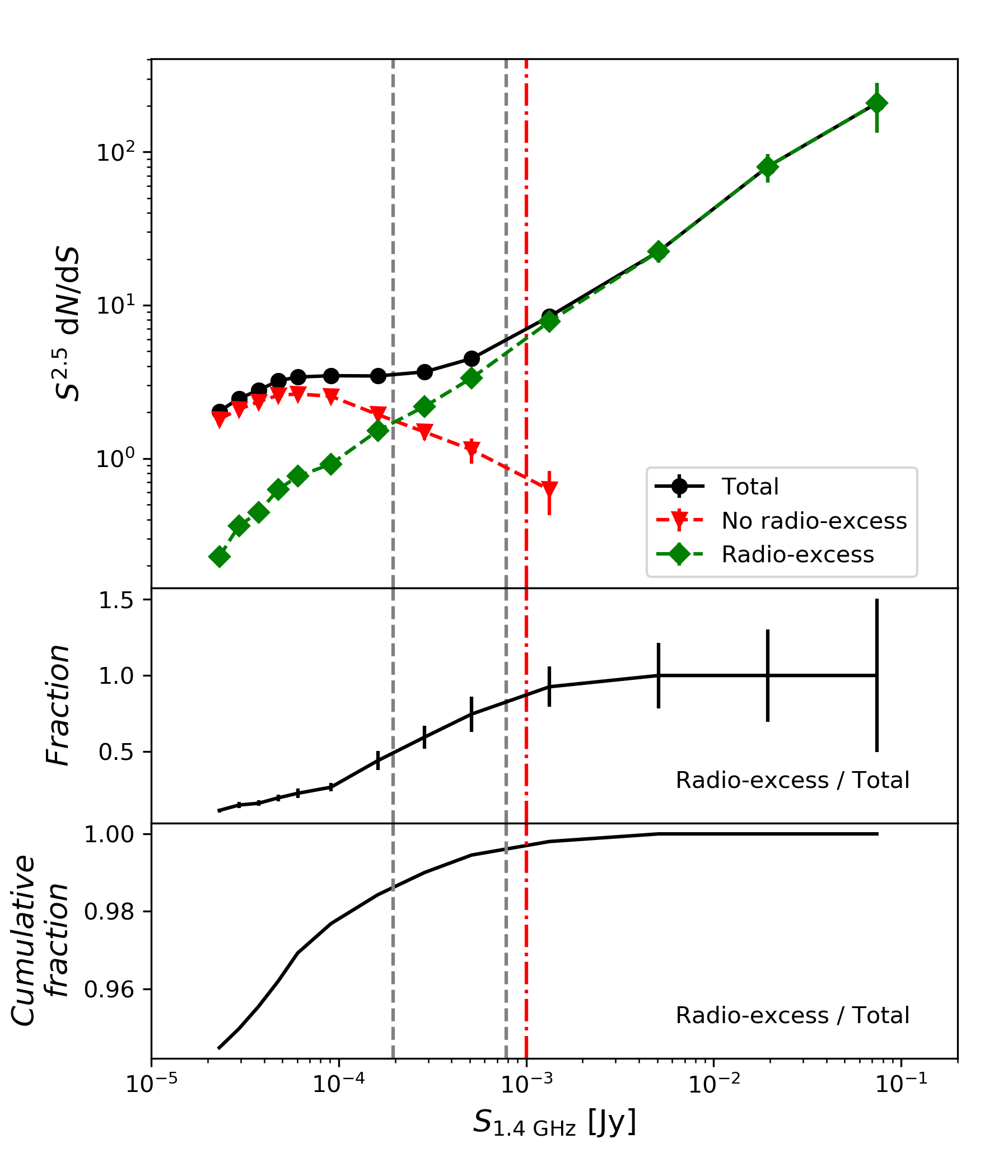}
\caption{Top panel: Euclidean-normalized and completeness-corrected source counts for different galaxy populations at $1.4 \ \mathrm{GHz}$ reproduced from \citet{Smolcic2017}, as described in the text (symbols indicated in the legend). The vertical gray lines correspond to the $7\sigma$ detection limits of the inner and outer part of the XXL-North GMRT survey recalculated from $610 \ \mathrm{MHz}$ by presuming a power law for the radio emission (see Section \ref{sec:Int}) and a spectral index of $-0.7$. Middle panel: Fraction of the radio-excess population. Lower panel: Cumulative fraction of the radio-excess population summed from higher fluxes towards lower. The red dot-dashed line denotes the adopted flux threshold described in the text.}
\label{fig:AGNFrac}
\end{figure}

Since we were interested in studying the evolution of AGN, we needed to assess the fraction of galaxy populations that constitute our sample. In order to do this, we used the VLA-COSMOS $3 \ \mathrm{GHz}$ catalog described in detail in \citet{Smolcic2017}. The main assumption here is that the galaxy populations obtained from one survey are comparable to other surveys, neglecting the effects of cosmic variance. We focused on the radio-excess sources described in \citet{Smolcic2017}, as this criterion is a good tracer of all AGN in the radio regime. \citet{Smolcic2017} defined the radio-excess sources when their radio luminosity $L_{1.4 \ \mathrm{GHz}}$ exceeded an extracted star formation rate luminosity given by 
$\log (L_{1.4 \ \mathrm{GHz}} / \mathrm{SFR}_{\mathrm{IR}}) = 21.984 (1+z)^{0.013}$. The star formation rate $\mathrm{SFR}_{\mathrm{IR}}$ was obtained by SED fitting from the total IR emission as described in \citet{Delvecchio2017}. By plotting the source counts of sources with and without radio-excess (see Fig. \ref{fig:AGNFrac}), and calculating the fraction of radio-excess sources, we concluded that our sample consists mostly of AGN. It can be seen from the cumulative function given in Fig. \ref{fig:AGNFrac} that at $7 \sigma = 350 \ \mathrm{\mu Jy}$, which is the lowest detection limit of our survey, we still have a sample that consists of more than $98\%$ AGN. However, the differential fraction  in the middle panel of the figure shows that the fainter bins also contain star-forming galaxies (SFGs). In order to obtain a pure sample of AGN,  we limited our sample to sources with flux density of $S_{610 \ \mathrm{MHz}} > 1 \ \mathrm{m Jy}$. This threshold brought the number of sources down to $1266$ (see Table \ref{tab:1}).

\subsection{Luminosity functions computation}
\label{AGN_Cre}
For the creation of the luminosity functions we follow the procedure outlined in \citet{Novak2017}, which relies on calculating the maximum observable volume for each galaxy (see \citealt{Schmidt1968}, \citealt{Felten1976}, \citealt{Avni1980}, \citealt{Page2000} and \citealt{Yuan2013}). The creation of luminosity functions is biased to the radio survey detection limit, so we take into account that the more luminous sources are detectable over larger distances (\citealt{Page2000}). The value of the luminosity function in each luminosity and redshift bin $\Phi (L,z)$ was calculated as the sum of inverse maximum volumes $1/V_{Max,i}$. The uncertainty of the luminosity functions, $\sigma_{\Phi}$, was calculated assuming Gaussian statistics (\citealt{Marshall1985}, \citealt{Boyle1988}, \citealt{Page2000}, \citealt{Novak2017}) and is not applicable to bins with very few sources. When the number of sources was lower than $10$, we used the tabulated errors determined by \citet{Gehrels1986}.

To calculate $V_{Max,i}$, we divided the complete sample into redshift subsets. For each subset the $V_{Max,i}$ estimation was performed independently. If the maximum volume exceeded the volume defined by the redshift bin, then the upper limit of the bin was used to determine $V_{Max}$. The spectral index was set to a fixed value of $\alpha = - 0.7$, which is consistent with the mean value calculated in Paper XXIX. A fixed spectral index simplifies the bin selection process described in Section \ref{BinSelection} by introducing a clear limit in the luminosity-redshift relationship. Furthermore, the luminosity functions were scaled to the area of observations $A$ by dividing it by the area of the celestial sphere as $A/41 \ 253\ \mathrm{deg^2}$.

\subsection{Corrections}
\label{LFCorr}
During the calculation of $V_{Max}$ a few corrections must be performed. The first correction accounts for the IRAC dataset depth. This correction is a function of redshift $C_{IRAC}(z)$ and accounts for the sources lost during the cross-matching of the radio catalog with the IRAC data. It was already discussed in Sect. \ref{IRAC_Corr}. A second correction must be applied due to the presence of noise in the observed radio map. Since the local value of noise differs from the mean noise (used to select the sources with $S/N > 7$), it follows that the true flux densities of some sources can fall below the detection limit. To assess the resulting incompleteness, and the corresponding correction $C_{Radio}(S_{610\ \mathrm{MHz}})$, we used observations from a deeper survey, namely the VLA-COSMOS $3\ \mathrm{GHz}$ Large Project (see \citealt{Smolcic2017a}) and compared the source counts. The detailed account of this correction can be found in Paper XXIX. 

The total correction applied to the GMRT-XXL radio data matched to IRAC counterparts was calculated following \citet{Novak2017} as the product of the above-mentioned corrections,
\begin{align} 
C_{Total} = C_{IRAC}(z) \times C_{Radio}(S_{610 \ \mathrm{MHz}}),
\end{align}
where $C_{IRAC}(z)$ is shown in Fig. \ref{fig:CosmosCut}, while $C_{Radio}(S_{610 \ \mathrm{MHz}})$ can be seen in Fig 13. of Paper XXIX. The calculated values of $V_{Max}$ were then multiplied by this number. After imposing a redshift and flux density threshold ($z\leq 2.1$, $S_{610\ \mathrm{MHz}}>1 \ \mathrm{m Jy}$) described in Sects. \ref{IRAC_Corr} and \ref{AGN_Fract}, and merging the catalogs for the inner and outer parts of the XXL-North field, we were left with a catalog of $1150$ sources, which was used in the creation of the luminosity functions (see Table \ref{tab:1}).

\begin{table}[]
\caption{Number of sources, and corresponding area, after each step performed during the analysis and luminosity function creation, as described in the text. The steps are performed progressively, i.e., each step also includes  the previous ones. }
\centering
\begin{tabular}{|p{2.5cm}||p{1.5cm}|p{1.5cm}|p{2cm}|}
     \hline
     \multicolumn{4}{|c|}{Outer part of the XXL-North field} \\
     \hline
      Step & Area$[\mathrm{deg}^2]$ & N(Radio) & N(Matched) \\
     \hline
     Complete catalog & 18.5 & 4615 & (...)  \\
     IRAC coverage & 16.7 & 4241 & 2954  \\
     Far from edge & 14.2 & 3499 & 2467  \\
     $S_{610 \ \mathrm{MHz}} > 1 \ \mathrm{m Jy}$ & 14.2 & 1605 & 948  \\
     $z\leq 2.1$ & 14.2 & (...)  & 855  \\
 \hline
\end{tabular}
\vspace{7mm}

\begin{tabular}{|p{2.5cm}||p{1.5cm}|p{1.5cm}|p{2cm}|}
     \hline
     \multicolumn{4}{|c|}{Inner part of the XXL-North field} \\
     \hline
      Step & Area$[\mathrm{deg}^2]$ & N(Radio) & N(Matched) \\
     \hline
     Complete catalog & 11.9 & 819 & (...)  \\
     IRAC coverage & 8.0 & 596 & 382  \\
     Far from edge & 6.3 & 477 & 318  \\
     $S_{610 \ \mathrm{MHz}} > 1 \ \mathrm{m Jy}$ & 6.3 & 477 & 318  \\
     $z\leq 2.1$ & 6.3 & (...)  & 295  \\
 \hline
\end{tabular}
\label{tab:1}
\end{table}

\subsection{Bin selection}
\label{BinSelection}
\begin{figure}
\includegraphics[width=0.5\textwidth]{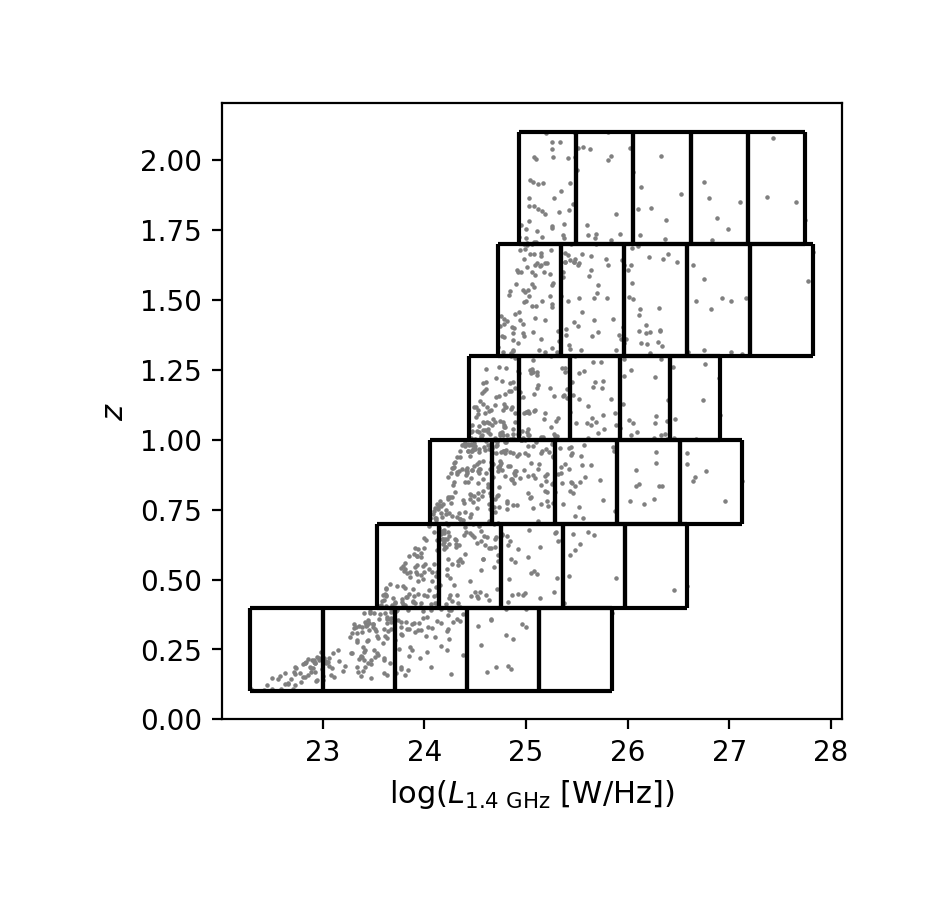}
\caption{Visual representation of the bins used in the creation of the luminosity functions. The gray dots represent the sources. The black lines correspond to the bin limits in redshift and luminosity. The absence of low-luminosity bins with only few sources is clearly visible. On the high-luminosity end the number of sources per bin decreases, but this effect is a consequence of the intrinsic lower density of high-luminosity sources and cannot be easily corrected.}
\label{fig:Lz}
\end{figure}

A rather subtle effect was observed by \citet{Yuan2013} which can lead to potential systematic errors. In short, if the redshift and luminosity bins are selected arbitrarily, the detection limit of the survey can introduce an unphysical bias. More specifically, because of the detection limit, there will be low-luminosity bins which enclose a very small number of sources. Apart from the problems associated with small number statistics, $[V_{Max} \ \Delta \log L ]$, present in the calculation of the luminosity functions, leads to an underestimation of $\Phi(L,z)$. As proposed by \citealt{Yuan2013}, a simple way to reduce this effect is to choose the luminosity bins so that they start from the value determined by the detection limit. A visual representation of this can be seen by looking at the luminosity-redshift plot shown in Figure \ref{fig:Lz}. For each redshift bin, the luminosity bins are set to start from the line defined by the detection limit, a method that ensures that no low-luminosity bin contains  very few objects.

\begin{figure*}
\centering
\includegraphics[width=0.9\textwidth]{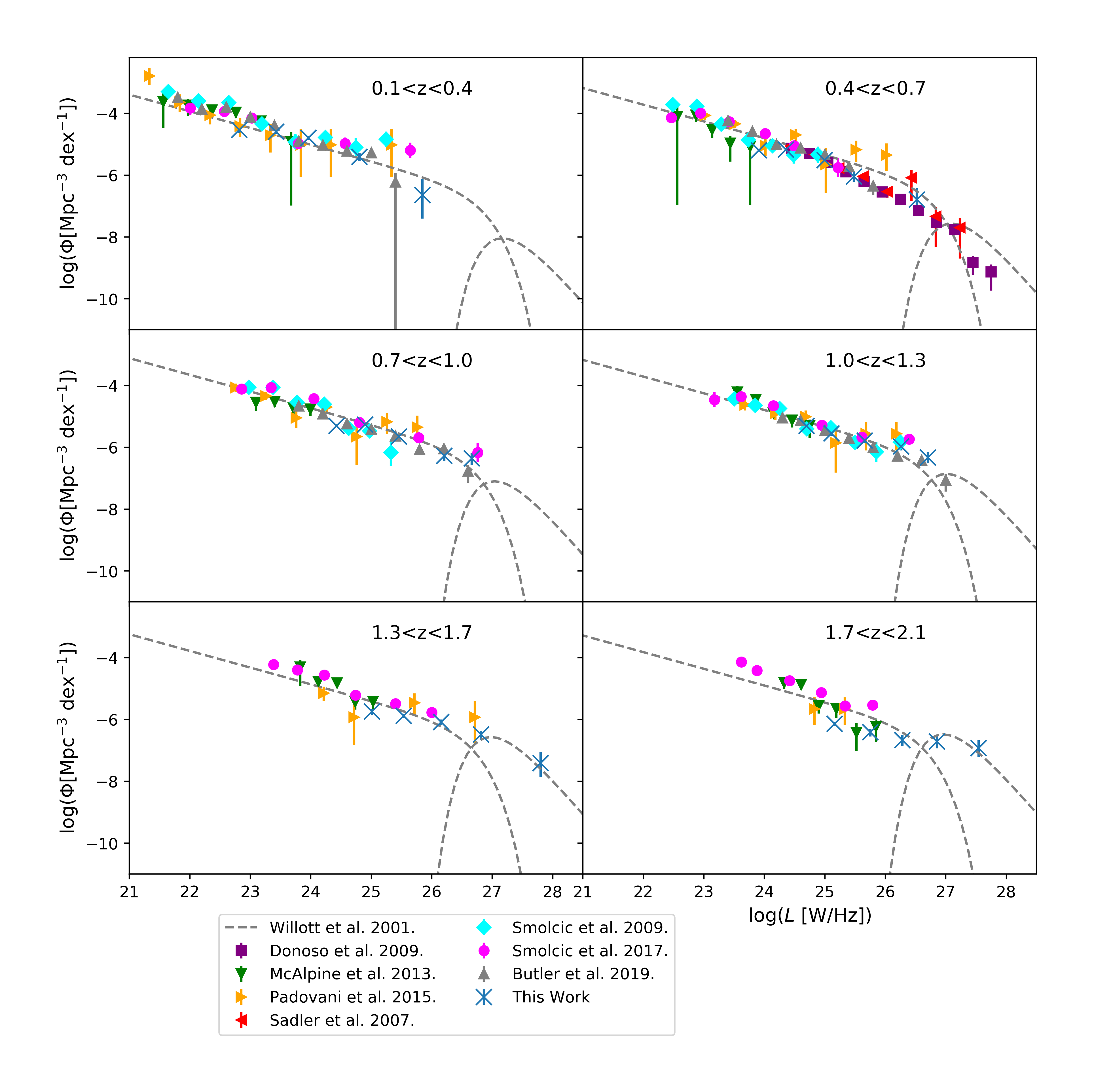}
\caption{Luminosity functions of this work along with previous ones at $1.4 \ \mathrm{GHz,}$ as denoted in the legend. The dashed lines represent the bimodal model discussed in the text. This model consists of a high- and a low-luminosity end with different functional dependencies. It can be seen that at higher luminosities, the high-luminosity end of the model traces the data points well.}
\centering
\label{fig:LF}
\end{figure*}

\section{Results}
\label{sec:LFRes}
The resulting radio luminosity functions from $z=0.1$ up to $z=2.1$ are shown in Figure \ref{fig:LF}. Since we made a comparison with luminosity functions at rest-frame $1.4 \ \mathrm{GHz}$, we also created the radio luminosity functions at this frequency (by assuming a power-law flux spectrum, and a spectral index of $-0.7$). The sampled luminosities depend on the redshift bin, with the maximum luminosities approaching $L_{1.4 \ \mathrm{GHz}} = 10^{28} \ \mathrm{W/Hz}$, as can be seen from the figure. 

\subsection{Comparison with the literature}
In Fig. \ref{fig:LF} we compare our data to a number of other studies. We show the radio luminosity functions by \citet{Sadler2007} derived from the volume-limited sample of $391$ radio galaxies with the optical  spectra from the 2dF-SDSS (Sloan Digital Sky Survey) LRG (Luminous Red Galaxy) and QSO (quasi-stellar object) surveys (2SLAQ; \citealt{Cannon2006}) with Faint Images of the Radio Sky at Twenty centimetres (FIRST; \citealt{Becker1995}) and  the NVSS radio coverage over redshifts $0.4 < z < 0.7$. They found that the radio emission in these sources most likely arises from AGN acitivity, rather than star formation. 

The radio luminosity functions by \citet{Donoso2009} were derived from the sample of $14453$ radio-loud (RL) AGN at redshifts $0.4<z<0.8$ detected at $1.4 \ \mathrm{GHz}$ with NVSS and FIRST radio surveys, previously cross-matched with the MegaZ-luminous red galaxy (MegaZ-LRG) catalog (\citealt{Collister2007}),  derived from the Sloan Digital Sky Survey. The large number of sources resulted in the notably smaller error bars of these functions, compared to previous studies in this redshift range (e.g., \citealt{Sadler2007}).

The luminosity functions by \citet{Butler2019} (hereafter XXL Paper XXXVI) come from a sample of $6287$ sources from the $2.1 \ \mathrm{GHz}$ observations of the XXL-South field, matched with the corresponding multi-wavelength catalog (Paper XVIII, Paper XXVI, Paper XXXVI). We show here the  RL AGN, classified by their radio excess (see \citealt{Butler2018}, XXL Paper XXXI, for classification details), which span redshifts up to $z = 1.3$. Our results are in good agreement with these surveys. 

The luminosity functions by \citet{McAlpine2013} come from a survey of VIDEO-XMM3 field at $1.4 \ \mathrm{GHz}$, aiming to investigate the evolution of faint radio sources (up to $100\ \mathrm{\mu Jy}$), up to a redshift of $z \approx 2.5$. The radio observations were performed with the VLA, with the photometric redshifts coming from the cross-matching with the Visible and Infrared Survey Telescope for Astronomy Deep Extragalactic Observations (VIDEO; \citealt{Jarvis2013}) and Canada–France–Hawaii Telescope Legacy Survey (CFHTLS; \citealt{Ilbert2006}). The sample consist of both  SFGs and AGN. The agreement between the AGN-related radio luminosity functions is good, although the overlap in luminosities in not large since the luminosity functions by \citet{McAlpine2013} mostly sample lower luminosities. 

The radio luminosity functions by \citet{Padovani2015} were derived from the sample of $680$ sources detected and identified within Extended Chandra Deep Field South (E-CDFS; \citealt{Bonzini2012}, \citealt{Miller2013}) using the $1.4 \ \mathrm{GHz}$ radio data observed with the VLA and cross-matched with the available multi-wavelength data. They probe the faint radio sky down to $\mathrm{\mu Jy}$ sources. Within the error bars, the agreement of their RL AGN luminosity functions with ours is good although the uncertainties become large at higher luminosities. This is due to a somewhat smaller area ($\approx 0.32 \ \mathrm{deg.}^2$) analyzed by \citet{Padovani2015}.

The luminosity functions by \citet{Smolcic2009} come from a sample of around $600$ AGN detected within the $1.4 \ \mathrm{GHz}$ VLA–COSMOS survey (\citealt{Schinnerer2007}). The luminosity functions consist of low-luminosity ($L_{1.4 \ \mathrm{GHz}} \leq 5 \times 10^{25} \ \mathrm{W \ Hz^{-1}}$) radio AGN at intermediate redshifts up to $z \approx 1.3$. The agreement with our data is good.

The luminosity functions by \citet{Smolcic2017c} come from the VLA-COSMOS $3 \ \mathrm{GHz}$ Large Project mentioned in Sects. \ref{IRAC_Corr} and \ref{AGN_Fract}, together with the VLA-COSMOS $1.4 \ \mathrm{GHz}$ Large and Deep Projects (\citealt{Schinnerer2004}, \citealt{Schinnerer2007}, \citealt{Schinnerer2010}). The sample consists of over $1800$ radio AGN, up to redshifts of $z \approx 5$. The large depth of the survey ensured the small uncertainties even at high redshifts. The agreement with our data is good, although the overlap in luminosities becomes smaller at higher redshifts, given the difference in observed areas and depth of the surveys. Apart from the luminosity functions we also show the model by \citet{Willott2001} denoted by a gray dashed line. For details on this model see the discussion in Sect. \ref{sec:Dis}.

\section{Discussion}
\label{sec:Dis}
\subsection{Cosmic evolution of the radio AGN population}
Here we have derived the the rest-frame $1.4\ \mathrm{GHz}$ radio luminosity functions for radio AGN out to $z\approx 2.1$ using the $610 \ \mathrm{MHz}$ GMRT survey comprising of intermediate luminosity AGN ($ 23 \lesssim \log(L_{1.4 \ \mathrm{GHz}}[\mathrm{W/Hz}]) \lesssim 28$) due to its $\sim 25  \ \mathrm{deg}^2$ surface area. Such luminosities are missed by deep radio surveys such as COSMOS/VIDEO which usually cover much smaller areas. In Fig. \ref{fig:LF} we compared our values and the literature $1.4 \ \mathrm{GHz}$ luminosity functions for radio AGN with the model presented by \citet{Willott2001}. The authors obtained their sample from shallow but large area surveys, namely the 7C Redshift Survey (7CRS) and the 3CRR and 6CE surveys at brighter luminosities (see Fig. 1 in \citealt{Willott2001} for details on luminosity range) at low frequencies ($151$ for the 3CRR and $178 \ \mathrm{MHz}$ for the other), which yielded $356$ sources. The radio luminosity functions were modeled using a two-population model that assumes different shapes and evolution properties for the high- and low-luminosity ends of the sample. We concentrate here on ``Model C'' described by \citet{Willott2001}. The low-luminosity end was modeled by a Schechter function (see relation 5 in \citealt{Willott2001}), while the high-luminosity end was modeled by a similar function (a Schechter function with inverted functional dependency for higher and lower luminosities; see relation 6 in \citealt{Willott2001}). The evolution of the low-luminosity end was modeled as a pure density evolution up to $z \approx 0.7$ (see Table 1 from \citealt{Willott2001}), after which the evolution ceases. The high-luminosity evolution was modeled by an asymmetric Gaussian function in redshift. The one-tailed Gaussian rise to redshift $z \approx 2$ was allowed to have a different width than the one-tailed decline at higher redshifts (see Table 1 from \citealt{Willott2001} for exact values).  

This evolution modeled by \citet{Willott2001} is consistent with the luminosity functions from this study. The standard Schechter form of the local luminosity function did not describe the data points at the high-luminosity end properly, given an excess in volume densities at high redhift and high luminosities. Therefore, following \citealt{Smolcic2009}, we compared our luminosity functions to the model (Model C) and the evolution parameters from \citet{Willott2001}, but recalculated to our cosmology and the frequency of $1.4 \ \mathrm{GHz}$. It can be seen, however, that the luminosity function model, determined by \citet{Willott2001}, follows our data points well, which could suggest that the high-luminosity population of AGN evolves more rapidly than the low-luminosity end. The discrepancies at lower luminosities and high redshifts are known issues with the model (as discussed in \citealt{Willott2001}).

Furthermore, it is widely reported in the literature (\citealt{Willott2001}, \citealt{Waddington2001}, \citealt{Clewley2004}, \citealt{McAlpine2013}, \citealt{Rigby2011}, \citealt{McAlpine2013}, \citealt{Rigby2015}) that a difference exists between the evolution of high- and low-luminosity sources. When the sample is divided into high- and low-luminosity sources, the comparison is straightforward. These papers (\citealt{Waddington2001}, \citealt{Clewley2004}, \citealt{Sadler2007}, \citealt{Smolcic2009}, \citealt{Donoso2009}, \citealt{Padovani2017}) find a difference in the evolution of high- and low-luminosity sources, where the high-luminosity sources are the ones that evolve faster. We also mention here  \citet{McAlpineJarvis2011} since the radio data comes from the XMM-LSS field mentioned in Section \ref{sec:radio} (\citealt{Tasse2007}). Their results are also consistent with this work, i.e., a bimodal evolution is found for high- and low-luminosity sources. Even in cases when the classification of sources is not identical to ours, the results lean towards a bimodal evolution. Whether the population is divided into RL and radio-quiet (RQ) AGN (e.g., \citealt{Padovani2015}) or into HERGs and LERGs (e.g., \citealt{Pracy2016}, Paper XXXVI), the evolutionary trends are still consistent. In other words, even if the classification is not exactly one-to-one the data always seem to lean towards a bimodal evolution where the sources with higher luminosities evolve faster. This trend can be explained by invoking the bimodality in the underlying physical picture, as described in the next subsection. We note  again that the luminosity functions presented here simultaneously reach high luminosities ($  \log(L_{1.4 \ \mathrm{GHz}}[\mathrm{W/Hz}])  \approx 28$) and redshifts ($z \approx 2.1$).

\subsection{Physical interpretation}
The results, from  this study and from the literature, can be explained by an underlying physical picture of the AGN outlined in the introduction. Evidence exists for  two physically different AGN populations (see \citealt{Hardcastle2007}, \citealt{Heckman_Best2014}, \citealt{Smolcic2017a},
\citealt{Padovani2017}): the radiatively efficient population and the radiatively inefficient population, the main difference between the two populations being their mode of accretion (\citealt{Hardcastle2007}, \citealt{Heckman_Best2014}, \citealt{Narayan1998}, \citealt{Shakura1973}). The radiatively efficient population, fueled by the cold intergalactic medium, exhibits higher Eddington ratios and evolves faster. In the literature they correspond to HERGs, while in our study they correspond to the high-luminosity end. The radiatively inefficient population, fueled by the hot intergalactic medium, evolves less rapidly and radiates at lower Eddington limits. This population corresponds to LERGs, or the low-luminosity end of our sample. In summary, the results presented within this work are consistent with both the earlier findings from the literature and the currently accepted physical interpretation of these findings.

\

\section{Summary and conclusion}
\label{sec:Sum}
We presented a study of AGN using the radio data from the GMRT radio telescope in the XXL-North field, and the corresponding multi-wavelength data (see Sect. \ref{MultiCat}), we were able to obtain a large sample of sources with photometric redshifts. A very careful cross-matching using the likelihood ratio method resulted in a catalog of $1150$ sources, whose radio emission is dominated by AGN processes, covering a rather large area of the luminosity--redshift plot ($z\leq 2.1$, $S_{610\ \mathrm{MHz}}>1 \ \mathrm{m Jy}$). We constructed the radio luminosity functions at $1.4 \ \mathrm{GHz}$ using the standard $V_{Max}$ method and examined their evolution and how it changes for low-luminosity and high-luminosity populations. The luminosity functions are in agreement with a double-population model from \citet{Willott2001}, supporting bimodal evolution found across the literature. The advantage of this survey was  that we could simultaneously reach redshifts of up to $z \approx 2.1$ and luminosities up to $\log(L_{1.4 \ \mathrm{GHz}}[\mathrm{W/Hz}]) \approx 28$, owing to the large area and depth of the observed field. 

\begin{acknowledgements}
VS, MN, LC, KT acknowledge the European Union's Seventh Framework programme under grant agreement 337595 (CoSMass). BS and VS acknowledge the financial support by the Croatian Science Foundation for project IP-2018-01-2889 (LowFreqCRO). XXL is an international project based around an XMM Very Large Programme surveying two $25 \ \mathrm{deg}^2$ extragalactic fields at a depth of $\sim 6 \cdot 10^{-15}\ \mathrm{erg}\ \mathrm{cm}^{-2} \mathrm{s}^{-1}$ in the $[0.5-2]\ \mathrm{keV}$ band for point-like sources. The XXL website is \url{http://irfu.cea.fr/xxl}. Multi-band information and spectroscopic follow-up of the X-ray sources are obtained through a number of survey programs, summarized at \url{http://xxlmultiwave.pbworks.com/}. The Saclay group acknowledges long-term support from the Centre National d'Etudes Spatiales (CNES).
\end{acknowledgements}

\bibliographystyle{aa}
\bibliography{Refs}

\end{document}